\documentclass[twocolumn,aip,reprint,unsortedaddress,amsmath,amssymb,floatfix]{revtex4-2}

\usepackage{amssymb,amsmath,graphicx,bm,textcomp,color,commath,bm}

\newcommand{\jerome}{}

\def \bfr {{\bf r}}
\def \bfk {{\bf k}}

\def \bfu {{\bf u}}
\def \bfe {{\bf e}}

\newcommand\doverline[1]{{\setbox0=\hbox{$\overline{#1}$}\ht0=\dimexpr\ht0-.15ex\relax
  \overline{\copy0}%
}}

\begin{document}

\title{Probing non-affine expansion with light scattering}

\author{Alesya Mikhailovskaya }\affiliation{Univ Rennes, CNRS, IPR (Institut de Physique de Rennes) - UMR 6251, F-35000 Rennes, France}
\author{Julien Fade}\affiliation{Univ Rennes, CNRS, Institut FOTON - UMR 6082, F-35000 Rennes, France}
\author{J\'er\^ome Crassous} \affiliation{Univ Rennes, CNRS, IPR (Institut de Physique de Rennes) - UMR 6251, F-35000 Rennes, France}
\email{jerome.crassous@univ-rennes1.fr}

\date{\today}

\begin{abstract}
	In disordered materials under mechanical stress, the induced deformation can deviate from the affine one even in the elastic regime. The non-affine contribution was observed and characterized in numerical simulations for various systems and reported experimentally in colloidal gels. However, low amplitude of non-affinity and its local character makes the experimental study challenging. We present a novel method based on the phase compensation of the wave scattered from a thermally dilated amorphous material using fine wavelength tuning of the optical probe beam. Using a glass frit as a sample, we ensure complete reversibility of the material deformation while experimental observations enable us to confirm the occurrence of non-affinity in the elastic regime. We develop a model for the coupled effect of the thermal expansion/contraction of the material and the dilatation of the incident wavelength which allows us to estimate the magnitude of the non-affine displacement and the spatial extent of its correlation domain.
\end{abstract}

\maketitle

\section{Introduction}
A slightly deformed solid behaves elastically: when a mechanical stress is applied to it, its shape gets distorted, and when the stress is removed, the solid recovers its original shape. The applied mechanical stress slightly changes the inter-atomic distances in the solid, which hence affects the energy of the system. The link between the geometry of the deformation and the energy of the material is usually presented in terms of affine deformation~\cite{slaughte.book}. The energy of the material may then be expressed as a function of  the so-called deformation tensor~\cite{slaughte.book}. However, except for perfect crystalline structures, such affine deformation creates  extra forces in the system, which, in return, give rise to an additional deformation in the system~\cite{alexander.1998,didonna.2005}. Understanding and characterizing such non-affine deformations is an important challenge, since the deviation between the actual deformation and the affine model may strongly impact the mechanical properties of the material. These effects have been observed in numerical simulations for systems of weakly connected disordered interacting particles close to isostaticity where important non-affine deformations take place~\cite{wyart.2011,ohern.2003,vanHecke.2009,zaccone.2011}. Similar influence has also been numerically confirmed for various amorphous systems~\cite{langer.1997,tanguy.2002,didonna.2005,aime.2018,basu.2011}.

From an experimental point of view, non-affine displacements have been reported to arise in the plastic regime~\cite{debregeas.2001,pomella.2019}, but the direct observation of such phenomenon in the elastic regime is quite intricate with only a few contributions reported so far in the literature. The difficulty arises from the required sensitivity of the measurements: except very close to isostaticity, non-affine deformations are expected to be of lower magnitude than the affine ones, which themselves must be small in the elastic regime. Moreover, deviations from the affinity occurs locally, and therefore, spatially resolved characterization of the displacement field is needed. It is likely the reason why such non-affine elastic deformations have been probed so far by studying acoustic properties of amorphous solids. In this case, it has been shown that the scattering of sound in amorphous solids~\cite{monaco.2009} may be explained in terms of non-affine deformations~\cite{caroli.2019}.

\begin{figure}[htbp]
\centering
\includegraphics[width=.99\columnwidth]{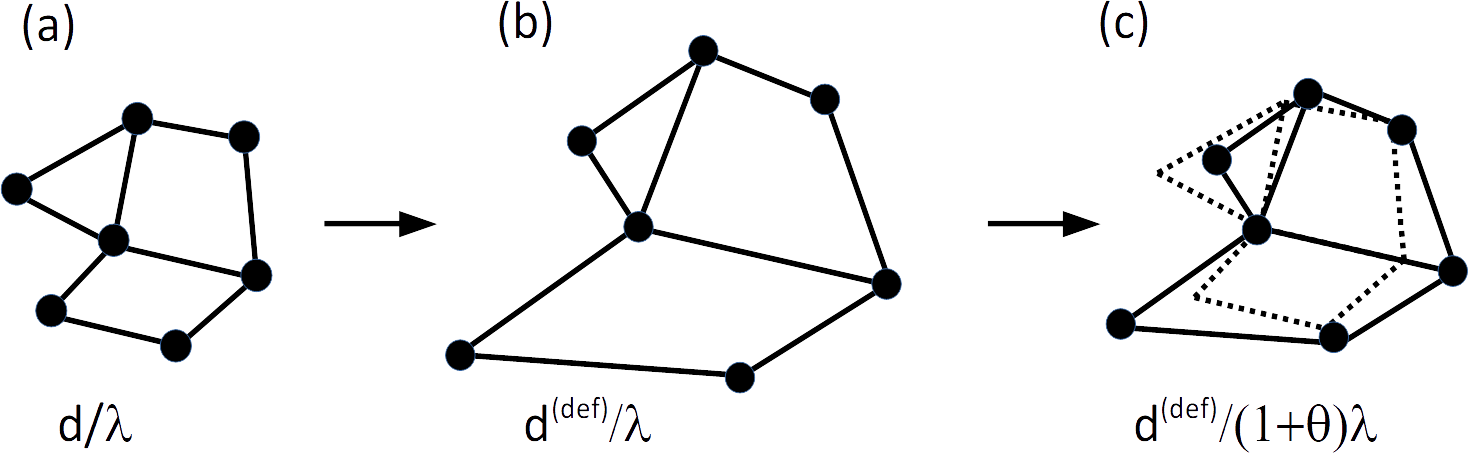}
\caption{(a) A disordered material viewed as a schematic 2D-set of connected vertices. (b) The material after dilatation. (c) Plain lines and vertices: deformed material after isotropic contraction by a factor $(1+\theta)$. Dotted lines: original material structure.}
\label{fig_principle}
\end{figure}

In this work, we propose another experimental approach, based on optical laser light scattering, where non-affine deformations may be evidenced in a disordered material. For that purpose, a solid sample considered as a dense packing of connected particles is subjected to an isotropic thermal dilatation, as schematically illustrated in~Fig.~\ref{fig_principle}.a and \ref{fig_principle}.b.  The principle of the experiment consists in studying the coherent scattering from this disordered material when it is illuminated with a coherent optical beam of wavelength $\lambda$. Upon dilatation of the material with a slight heating, the characteristic inter-particle distance $d$ is changed into a somewhat larger distance $d^{(def)}$, as depicted in Fig.~\ref{fig_principle}.b. If the wavelength of the probing light is now dilated by a factor $(1+\theta)$, and if $\theta$ is such that $d^{(def)}=(1+\theta)d$ in the system, then the ratio $d/\lambda$ should remain unchanged after these successive operations. However, if the mechanical deformation in the material is not perfectly affine, the material expansion is not homogeneous and the deformed network cannot be matched with the original one, as sketched in Fig.~\ref{fig_principle}.c. As a result, the above relation $d^{(def)}=(1+\theta)d$ is not simultaneously verified for all the inter-particle distances, and the ratio $d/\lambda$ does not take on a unique value in the deformed material. Thus, if some non-affinity takes place during those operations (heating + wavelength shift), the phase of the scattered optical wave given by the ratio of the propagation distance to the wavelength, is not conserved. As shown below, measuring the decorrelation function of an optical speckle intensity field makes it possible to retrieve quantitative information on non-affine deformations in the material. We showed in a previous study~\cite{Crassous.2009} that the phase shift due to thermal dilatation may be compensated only partially by wavelength expansion. However, the studied material represented a packing of non-connected glass beads that underwent an important irreversible displacement upon thermal expansion. Moreover, the range of wavelength variation was very narrow due to limitation of the laser source. In the present study, we consider a sintered packing of glass spheres (glass frit) which prevents such irreversible reorganisations. In addition, the used laser source allows to study material expansion of greater magnitude by ensuring a larger mode-hop-free spectral tuning range.

The paper is organized in the following way. In Section~\ref{sec.theory}, we discuss the behavior of waves scattered by a disordered material when both the material deformation and a variation of the incident wavelength occur. We derive the variation of the correlation function upon thermal expansion and wavelength shift. The experimental setup, the studied sample and the computation of the intensity correlation functions are described in Section~\ref{sec.experiment}. In  Section~\ref{sec.results}, we present a typical experiment combining thermal dilatation and wavelength shift of the laser, for which we compare the measured correlation functions with our model. We discuss in Section~\ref{sec.discussion} the values of the optical and mechanical parameters that we have determined from this experiment, in particular the amount of non-affine deformations observed, which is in agreement with the expected order of magnitude from theory or numerical simulations. A special care is taken in the Section~\ref{sec.discussion} to estimate the contribution of thermo-optic effects to the observed decorrelation, which is shown to be of insufficient magnitude to explain the obtained experimental results. Finally, we make a link between the magnitude of the observed non-affine deformations and their spatial extension.

\section{Theoretical model}\label{sec.theory}
\subsection{Optical phase variation along a deformed path}

\begin{figure}[htbp]
\centering
\includegraphics[width=0.9\columnwidth]{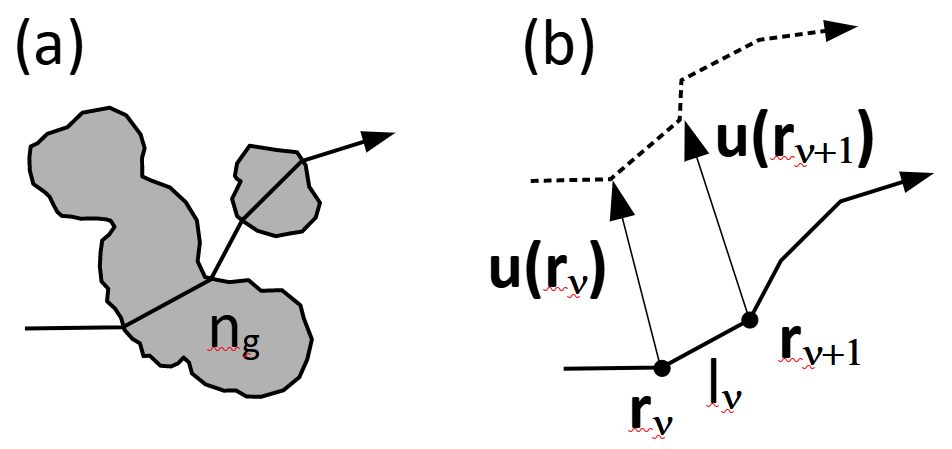}
\caption{(a) Schematic view of a part of an optical light ray inside a heterogeneous material composed of a medium of refractive index $n_g$ (shaded areas) and vacuum. (b) Decomposition of an optical light ray in the material into a succession of straight segments between spatial locations $\bfr_\nu$ and $\bfr_{\nu+1}$. Upon mechanical deformation of the material, the initial light ray (plain line) is modified into a new trajectory (dotted line), and the location $\bfr_\nu$ is displaced by a vector $\bfu(\bfr_\nu)$.}
\label{fig_path}
\end{figure}

We first consider one path of light inside a heterogeneous material composed of a medium of refractive index $n_g$ and  vacuum that is schematically represented in  Fig.~\ref{fig_path}.a. A path may be decomposed into a succession of $N$ linear segments separated by points $\bfr_\nu$, with $\nu$ an integer index $0\le\nu\le N-1$. Let $\nu$ denote the segment joining $\bfr_{\nu}$ to $\bfr_{\nu^+1}$, with $l_\nu=\norm{\bfr_{\nu+1}-\bfr_{\nu}}$ the length of this segment, and $\bfe_\nu=(\bfr_{\nu+1}-\bfr_{\nu})/l_\nu$ the unit vector along it, as illustrated in Fig.~\ref{fig_path}.b. We assign to each segment a variable $\eta_\nu$ which is $0$ if the segment is in vacuum, and $1$ otherwise, so that the segment lies in a medium of effective refractive index $n_\nu=1+(n_g-1)\eta_\nu$. The optical phase shift between the start and the end of such a path is therefore
\begin{equation}
\phi_N=k \sum_{\nu=0}^{N-1} n_\nu l_\nu,
\end{equation}
with $k=2\pi/\lambda$, and where $\lambda$ denotes the  wavelength in vacuum. When the material is deformed, the pathways of light are changed, and the variation of the optical phase shift reads:
\begin{equation}
\delta\phi_N=\sum_{\nu=0}^{N-1} \bigl[\delta k ~ l_\nu ~ n_\nu +k ~ \delta l_\nu ~ n_\nu + k ~ l_\nu ~ \delta n_\nu \bigr],\label{eq_dphiN}
\end{equation}
where $\delta k$, $\delta l_\nu$ and $\delta n_\nu$ are the variations of $k$, $l_\nu$ and $n_\nu$, and we suppose that perturbations are small: $\delta k \ll k$,  $\delta l_\nu \ll l_\nu$, $\delta n_\nu \ll n_\nu$. We study the phase variations resulting from a direct change of the wavelength and from the deformation of the material occurred with a change in temperature, and then:
\begin{subequations}
\begin{align}
\delta k&=-k ~ \bigl(\frac{\delta \lambda}{\lambda}\bigr) \label{eq_dk} \\
\delta n_\nu&=\eta_\nu ~ \bigl(\frac{\partial n_g}{\partial T}\bigr)\delta T \label{eq_dnu} \\
\delta l_\nu&=\bigl(\frac{\delta l_\nu}{\delta T}\bigr) ~ \delta T. \label{eq_dlnu}
\end{align}
\end{subequations}
{\jerome In the above equation \eqref{eq_dnu}, we have neglected dispersion effects which would include an additional term  $\eta_\nu \bigl(\partial n_g / \partial \lambda \bigr)\delta \lambda$. Indeed, the contribution of wavelength dispersion to $\delta(k n_\nu l_\nu)$ can be written as $k l_\nu \eta_\nu (\partial n_g / \partial \lambda) \delta \lambda$, which can be compared to $\delta k~n_\nu~l_\nu=-k~(\delta \lambda /\lambda)~n_\nu~l_\nu$. For such materials as glasses that are studied experimentally in this work we have $ \vert \frac{\partial n_g}{\partial \lambda}\vert  \ll 1/\lambda$. It follows that $k l_\nu \eta_\nu (\partial n_g / \partial \lambda) \delta \lambda \ll \delta k~n_\nu~l_\nu$. In conclusion, the phase variations due to dispersion effects can be considered as negligible  with respect to the} phase variations due to a change in $\delta k$.

\subsection{Geometry of the deformation}

Since the phase variation $\delta \phi_N$ depends on $\delta l_\nu$ through Eq.~\eqref{eq_dlnu}, we now describe the deformation of a material subjected to a thermal expansion. The segment $\nu$ located initially between the points $\bfr_{\nu}$ and $\bfr_{\nu^+1}$ now joins the points $\bfr_\nu+\bfu(\bfr_\nu)$ and $\bfr_{\nu+1}+\bfu(\bfr_{\nu+1})$ after deformation, where $\bfu(\bfr)$ is the displacement field. For small deformations such that $\norm{\bfu(\bfr_{\nu+1})-\bfu(\bfr_\nu)} \ll l_\nu$ the length variation is then:
\begin{align}
\delta l_\nu \simeq \bfe_\nu \cdot \bigl[\bfu(\bfr_{\nu+1})-\bfu(\bfr_\nu)\bigr]\label{eq_delta_lnu}
\end{align}

{\jerome We can decompose the differential displacement $\bfu(\bfr_{\nu+1})-\bfu(\bfr_\nu)$ as the sum of an affine dilatation and a deviation from an isotropic deformation:
\begin{align}
\bfu(\bfr_{\nu+1})-\bfu(\bfr_\nu)
&=\alpha ~ \delta T ~ (\bfr_{\nu+1}-\bfr_{\nu}) \nonumber \\
&+\bigl[\delta\bfu(\bfr_{\nu+1})-\delta\bfu(\bfr_\nu)\bigr],
\label{eq_dlnu_geo}
\end{align}
where $\alpha$ is the coefficient of linear expansion.  Using Eq.~\eqref{eq_dlnu_geo}, the equation \eqref{eq_delta_lnu} becomes:
\begin{equation}
\delta l_\nu \simeq \alpha ~ \delta T ~ l_\nu + \bfe_\nu \cdot \bigl[\delta \bfu(\bfr_{\nu+1})-\delta \bfu(\bfr_\nu)\bigr].
\end{equation}
The decomposition of Eq.~\eqref{eq_dlnu_geo} is not unique, and we must add an additional constraint to enforce uniqueness. For that purpose, we impose that the non-affine displacement is a part of the displacement which is not correlated with the relative locations of the points $\bfr_\nu$, and hence is uncorrelated with $\bfe_\nu$, such that:
\begin{equation}
 \sum_{\nu=0}^{N-1} \bfe_\nu \cdot \bigl[\delta\bfu(\bfr_{\nu+1})-\delta\bfu(\bfr_\nu)\bigr]=0,
\label{eq_constraint1}
\end{equation}
where $\sum_{\nu=0}^{N-1}$ (or $\sum_\nu$ for the sake of conciseness) represents the summation over all the segments of the considered light path. This constraint may also be written as:
\begin{equation}
 \sum_\nu \delta l_\nu=\alpha \, \delta T \sum_\nu l_\nu
\label{eq_constraint2}
\end{equation}
which means that $\alpha \,\delta T$ accounts for the relative increase of the total length of all paths inside the material.}

The deviations from affine isotropic deformations are expected to vary linearly with the affine contribution~\cite{alexander.1998}, and hence with $\delta T$, allowing us to reasonably write:
\begin{equation}
[\delta \bfu(\bfr_{\nu+1})-\delta \bfu(\bfr_\nu)\bigr]=\bm{\beta}_\nu ~ l_\nu ~ \delta T,
\label{eq_beta_nu}
\end{equation}
where we introduce the vector $\bm{\beta}_\nu$ which does not depend on $\delta T$, and which verifies $\sum_\nu \bm{\beta}_\nu \cdot \bfe_\nu=0$. The phase variation given in Eq.~\eqref{eq_dphiN} may then be written as:
\begin{equation}
\delta\phi_N=k \sum_{\nu=0}^{N-1}  l_\nu n_\nu
\bigl[-(\frac{\delta \lambda}{\lambda})+A_\nu \delta T\bigr],
\label{eq_dphiN_bis}
\end{equation}
with
\begin{equation}
A_\nu=\alpha +  (\frac{\eta_\nu}{n_\nu}) ~ (\frac{\partial n_g}{\partial T}) + \bfe_\nu \cdot \bm{\beta}_\nu .
\label{eq_Anu}
\end{equation}

\subsection{ Averaged phase variations}

We  now calculate the average value of $\big\langle \exp(j \delta \phi_N)\big\rangle _N$, assuming $\big\langle \cdot\big\rangle _N$ to be the average of the quantity $\cdot$ over all light paths involving $N$ segments. Since we consider a heterogeneous material which scatters strongly the light, we have $N \gg 1$, and by the central limit theorem, $\delta \phi_N$ can be considered as a Gaussian random variable~\cite{pine.1990b,pine.1993}, so that~\cite{dainty,goodman.book}:
\begin{align}
\Big\langle \exp(j \delta \phi_N)\Big\rangle _N&=\exp\Big(j \big\langle \delta \phi_ N\big\rangle _N\Bigr)\nonumber\\
&~\exp\Bigl(-\frac{\big\langle \delta \phi_N^2\big\rangle _N-\big\langle \delta \phi_N\big\rangle _N^2}{2}\Bigr).
\label{aa}
\end{align}
From Eq.~(\ref{eq_dphiN_bis}), the mean phase shift and its variance may be written as a combination of $\delta T$ and $\delta \lambda/\lambda$, i.e.,
\begin{align}
\langle \delta\phi_N\rangle _N & =N  k  \overline{n} \langle l_\nu\rangle  ~ \big(A \delta T -\frac{\delta \lambda}{\lambda}\big)
\label{eq_dphiN_mean}
\end{align}
and
\begin{align}
\frac{\langle \delta \phi_N^2\rangle _N-\langle \delta \phi_N\rangle _N^2}{2}&=N  k  \overline{n}\langle l_\nu\rangle  ~ \nonumber \\
\Bigl[ B ~ \big(A \delta T &-\frac{\delta \lambda}{\lambda}\big)^2+ C ~ (\delta T)^2 \Bigr],
\label{eq_dphiN_mean2}
\end{align}
where $A$, $B$, $C$ are coefficients whose calculi are given in the Appendix. {\jerome In the above expressions, and throughout the remainder of this article, we have introduced the following notation $\overline{a}$ for a "length-average" of any quantity $a_\nu$ which depends on segment index $\nu$: $\overline{a}=\big\langle a_\nu l_\nu\big\rangle /\langle l_\nu\rangle $. Similarly, we define a quadratic average as $\doverline{a}=\big\langle a_\nu l_\nu^2\big\rangle /\langle l_\nu\rangle ^2$. Since $\big\langle l_\nu^2\big\rangle \neq \big\langle l_\nu \big\rangle ^2$, beware that $\overline{l}=\langle l_\nu l_\nu\rangle /\langle l_\nu\rangle \neq \langle l_\nu\rangle $, and that for any constant $c$: $\doverline{c}=\langle c~l_\nu^2\rangle /\langle l_\nu\rangle ^2=c \langle l_\nu^2\rangle /\langle l_\nu\rangle ^2 \neq c$.}

\subsection{Correlation functions}

We now suppose that the system is illuminated by a light beam, and we collect the scattered light in a given experimental geometry.  We introduce $P(s)$ as the normalized distribution of path lengths and, therefore, the electric field autocorrelation function reads~\cite{pine.1990b,pine.1993}:
\begin{align}\label{eq:corr_func}
g_E=\int_s P(s) \big\langle \exp (j \delta \phi_s)\big\rangle  ds,
\end{align}
where $\delta \phi_s$ is the phase variation for a light path of length~$s$. Identifying $s$ with $N\langle l_\nu\rangle $ in the above expression, and using Eqs.~\eqref{aa}-\eqref{eq_dphiN_mean2}, we obtain:
\begin{align}
g_E(p)=\int_s P(s) \exp (p ~ s) ds,
\label{gE}
\end{align}
with
\begin{align}
p &=k \overline{n} \Big[
j\big(A \delta T -\frac{\delta \lambda}{\lambda}\big)\nonumber\\ &-B ~ \big(A \delta T -\frac{\delta \lambda}{\lambda}\big)^2-C ~ \big(\delta T)^2 \big) \Big]\label{p}
\end{align}
The quantity $g_E(p)$ is the Laplace transform of $P(s)$. For a plane-parallel slab of thickness $L$ in backscattering geometry, $P(s)$ may be calculated as~\cite{vellekoop.2005}:
\begin{equation}
g_E(p)= \frac{J(\sqrt{(3 p/l^*)+\alpha_a^2})}{J(\alpha_a)}
 \end{equation}
with:
\begin{equation}
J(\xi)=\frac{\sinh((L-z_0) \xi )+z_e \xi \cosh((L-z_0) \xi)}{(1+(z_e \xi)^2)\sinh(L \xi)+2 z_e\xi  \cosh(L \xi)},
\label{eq_J}
 \end{equation}
where $l^*$ is the transport mean free path of the light inside the material, $z_0$ is the depth at which the diffusing source is located, and $z_e$ is the so-called extrapolation length~\cite{vellekoop.2005}. Finally, $\alpha_a$ is a coefficient related to the light absorption inside the material. The intensity correlation function (that we access experimentally in the remainder of the article) is finally related to the electric field autocorrelation function $g_{E}$ using the Siegert relation~\cite{goodman.book}:
\begin{equation}
g_I=\vert g_E \vert^2.
\label{gI}
\end{equation}

\section{Experiment}\label{sec.experiment}

\subsection{Experimental setup}\label{sec.experiment.A}

\begin{figure}[htbp]
\centering
\includegraphics[width=0.7\columnwidth]{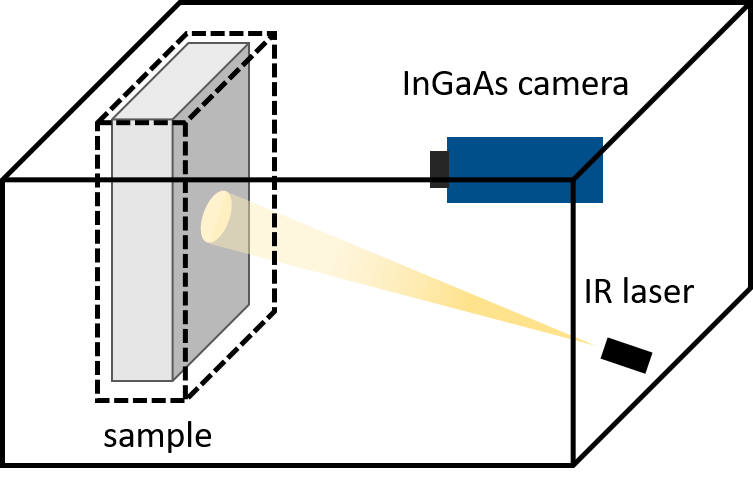}
\caption{Sketch of the experimental setup representing a DWS arrangement in backscattering configuration using a short-wave infrared laser with tunable wavelength as the illumination source, and an InGaAs camera as the detector. Two temperature controllers used respectively for the sample and for the enclosure are shown with dashed and solid lines, correspondingly.}
\label{setup}
\end{figure}

The experimental setup that we propose to use for sensing non-affine deformations in granular materials represents an arrangement of diffusing wave spectroscopy (DWS) in backscattering configuration and is schematically sketched in Fig.~\ref{setup}. The experiment being based on the fine measure of decorrelation of speckle interference intensity patterns, it was necessary to ensure the best thermal stability of the experiment. In order to reduce the relative displacement between the elements of the setup due to external thermal fluctuations, all the experiment (including laser, camera, regulated thermal cell) is placed into a temperature controlled environment that allows the whole setup to be kept at $22 \pm 0.2^{\circ}C$. The scattering material is placed into a thermo-regulated cell. The regulation of the cell is made with a PID controller (Stanford Research PTC10) which ensures a thermal stability of $\pm 3~mK$. In order to minimize thermal drift, the setup (including internal and external thermal regulated cells) was left to equilibrate for at least 24 hours.

The illumination source is a short-wave infrared (SWIR) fibred distributed feedback (DFB) laser (PowerSource 1905 LMI from Avanex, USA) combined with a laser diode controller (LDC-3744, ILX LightWave, USA). This device makes it possible to modulate the incident wavelength from $\lambda=1.5557~\mu$m to $1.55705~\mu$m  without mode hopping by finely tuning the temperature of the semiconductor laser chip. The spectral width of the laser diode is $<5$~MHz, giving a coherence length $> 60$~m, which is very large compared to the total path length inside the material. This means that the contrast of the speckle pattern does not depend on the spectral width of the laser.  The laser beam was used to shine the sample with a circular $6~mm$-diameter laser spot. Upon scattering of light in the diffusing material, the light backscattered out of the sample forms  a far-field speckle pattern, whose intensity fluctuations were recorded with an InGaAs SWIR camera (OWL 320, Raptor Photonics, Northern Ireland) that provides $320~$px $~ 256~$px resolution, with a pixel size of $30~\mu$m $~ 30~\mu$m. The average size of a speckle spot was measured to be approximately of $\simeq 3$ pixels per speckle.

The sample used in this experiment was a glass frit (grade P2) designed for filtration, and purchased from Bibby Scientific.  The thickness of the frit is $L=4.4$~mm, and the volume fraction of glass $\phi_g=0.648$ is measured by weighting of the frit. The P2 grade of the frit corresponds to a "maximum pore size" lying in the range $d_p=40-100~\mu$m. Microscopic inspection of the frit shows that it is composed of glass particles with a typical polydispersity of $\sim 3$. The frit is made of Borosilicate Pyrex Glass with the following physical characteristics~\cite{schott1,schott2}: the refractive index at $1.50~\mu$m is $n_g=1.456$, with a dispersion coefficient of $(\partial n_g/ \partial \lambda) = -1.2~ 10^{-5}$ nm$^{-1}$, and the coefficient of the linear thermal expansion is given as $\alpha=3.25~ 10^{-6}$~K$^{-1}$. Few references in the literature were found to assess the value of the thermo-optic coefficient of Pyrex glass, especially in the SWIR range. From references~\cite{ramachandran, peters}, the value of the $(\partial n_g/\partial T)$ can be expected to lie between $+5~ 10^{-6}$ and $+10^{-5}$ K$^{-1}$ for temperatures in the $20-100^\circ C$ range, and for a visible wavelength. According to experiments reported on other types of glass\cite{Harris}, the value of the thermo-optic coefficient tends to slightly decrease with an increase of the wavelength, but seems to remain of the same order of magnitude. A measure of the thermo-optic coefficient of a borosilicate glass with physical characteristics close to Pyrex has been reported at 1550~nm in~\cite{koike}, with a value close to $+9.10^{-6}$ K$^{-1}$. In the remainder of this article, we may thus retain an admissible range of $[+5~10^{-6}; +10^{-5}]$ K$^{-1}$ for the thermo-optic coefficient of Pyrex glass at 1550~nm. It can be noted here that the above numerical values allow us to validate the assumption made in the theoretical model to neglect dispersion effects. Indeed, with $(\delta \lambda)_{max} \sim 100$~pm and $(\delta T)_{max}\sim 15^\circ$ K in the experiment reported below, one has $(\delta T)_{max}.\bigl|\frac{\partial n_g}{\partial T}\big|\sim 70 ~ (\delta \lambda)_{max}.\bigl|\frac{\partial n_g}{\partial \lambda}\big|$.

\subsection{Correlation function calculus}\label{sec.experiment.B}

We obtain the normalized intensity correlation function between two images acquired at $(\lambda_1,T_1)$ and at $(\lambda_2,T_2)$ from the product of the recorded speckle intensity averaged on all pixels of the camera. However, we observed a drift of the speckle pattern with time even at rest because of the thermal expansion of the temperature regulation cell and due to the long duration of experiments. In order to correct this drift we proceed as follows: let $I_1(p)$ and $I_2(p)$ denote the raw intensities recorded at pixel $p$ in the image $1$ and $2$, respectively. We construct $I_{2,s}(p,\delta \bfr)$ as the intensity of the pixel $p$ when the image $2$ is shifted by a quantity of $\delta \bfr$ using a cubic-spline interpolation of intensity. Then, we define $g_I$ as:
\begin{equation}
g_I(1,2) = \mathrm{max} \Bigg\{\frac{\langle I_1 I_{2,s}\rangle - \langle I_1 \rangle\langle I_{2,s} \rangle}{\sqrt{\langle I_1^2\rangle - \langle I_1 \rangle ^2}\sqrt{\langle I_{2,s}^2\rangle - \langle I_{2,s} \rangle ^2}}\Bigg\}.
\label{eq_gI_exp}
\end{equation}
The maximization with respect to $\delta \bfr$ is obtained using a Powell maximization scheme~\cite{nr.book}. The shift $\delta \bfr$ corresponding to the maximal correlation is typically smaller than $2$ pixel size (i.e., $60 ~\mu$m) both in horizontal and vertical directions. However, the coherence area on the camera sensor being $\sim 100~\mu$m, neglecting this shift of the speckle pattern would diminish the correlation and would significantly bias the measurements.

\section{Experimental results}\label{sec.results}

\subsection{Temperature and wavelength control}

\begin{figure}[htbp]
\centering
\includegraphics[width=0.99\columnwidth]{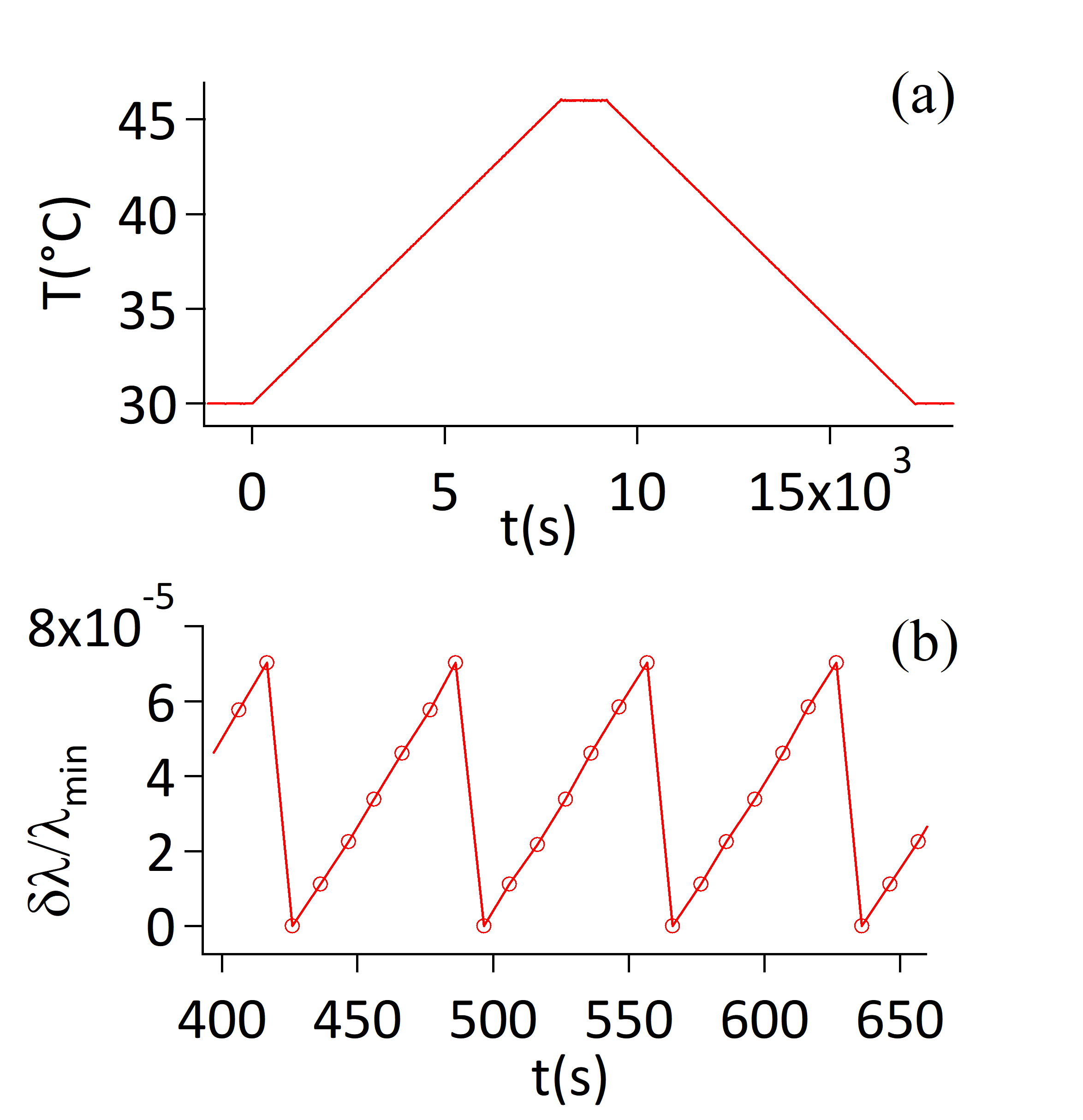}
\caption{(a) Evolution of the sample temperature during one experimental cycle. (b) Relative variation of the wavelength $\delta\lambda/\lambda_{min}$ with $\lambda_{min}$ the minimum wavelength, and $\delta\lambda=\lambda-\lambda_{min}$. The time origin is arbitrary.}
\label{fig_temp_lambda}
\end{figure}

The plots of Fig.~\ref{fig_temp_lambda} summarize a typical experiment. Initially, we keep the sample at the fixed temperature of $30^{\circ}C$ for a few minutes. Then we first increase the temperature with the constant rate of $2 ~ 10^{-3}~$K.s$^{-1}$ up to $46^{\circ}C$, hold it at $46^{\circ}C$ for several minutes, and finally we decrease the temperature back to $30^{\circ}C$ with the same rate. During this cycle, the laser wavelength is gradually increased from its minimal value $\lambda_{min}$ with steps of $18~$pm every $10~$s, and after $6$ successive steps the wavelength is set back to $\lambda_{min}$. The wavelength scanning cycle is thus approximately 280 times faster than the temperature cycle. Image recording on the camera is performed with a $10~$s period, therefore, during one full temperature cycle, we acquire approximately $1500$ images at different temperatures and wavelengths.

\begin{figure}[htbp]
\centering
\includegraphics[width=0.99\columnwidth]{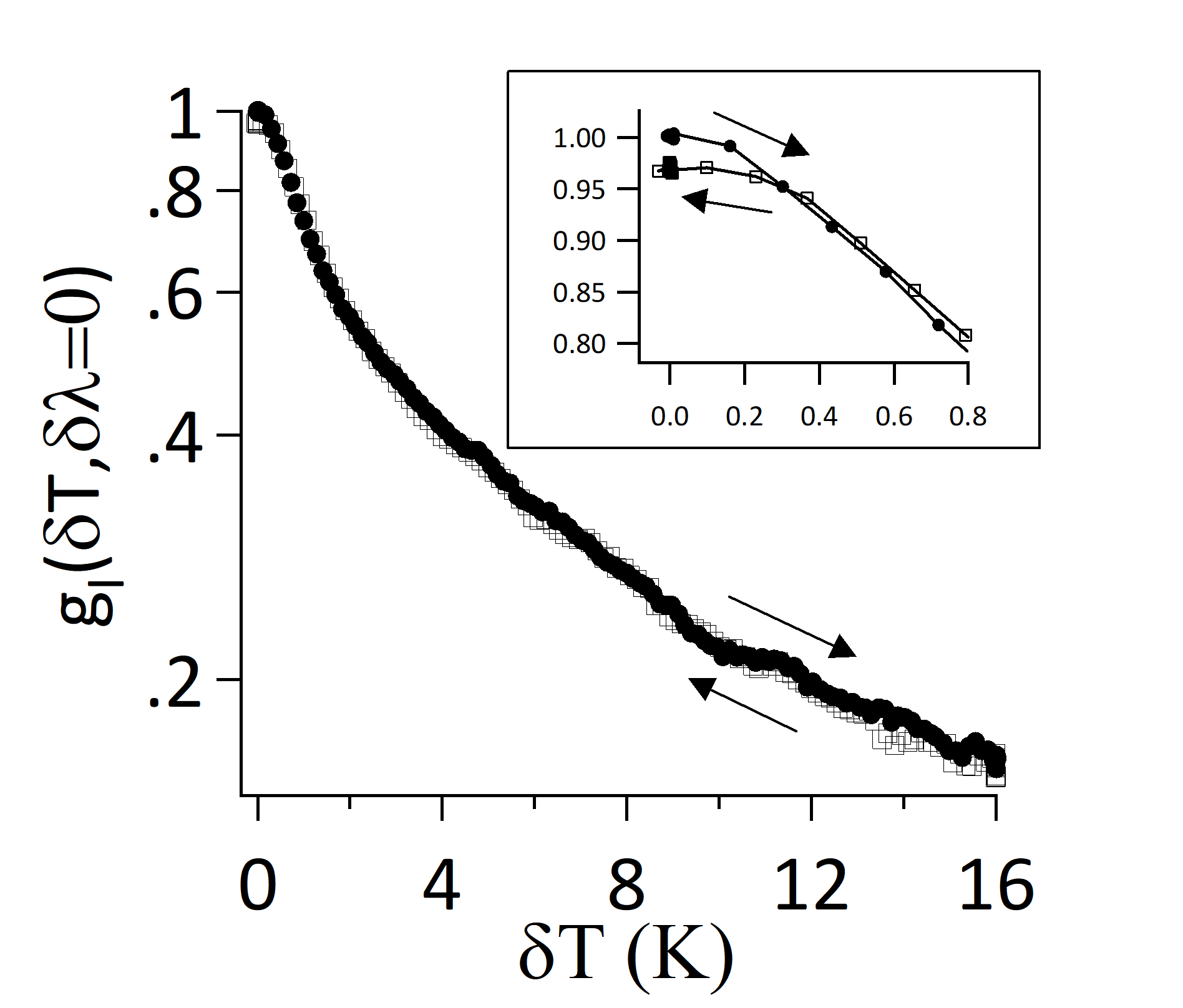}
\caption{Correlation function as a function of $\delta T$ when the temperature is first increased (filled circle) and then decreased (open square). Inset is a closeup for lower values of $\delta T$.}
\label{fig_reversibility}
\end{figure}

We confirm the reversibility of the sample deformation directly from the intensity correlation function obtained after one full experimental cycle comprising sequential dilatation and contraction of the sample. For this purpose, we correlate an image acquired before the increase of temperature with successive images recorded at the same wavelength. Figure \ref{fig_reversibility} shows the evolution of the intensity correlation function $g_I$ as a function of the temperature difference. We obtain $g_I(\delta T=0,\delta \lambda=0)\simeq 0.97$  after the temperature cycle indicating a good overall reversibility of the correlation function. We also examine the stability of the recorded signal by performing the experiment in which we keep constant both temperature and wavelength during a few hours. We observe a typical decrease of the correlation of about $0.01$ per hour. We attribute this small decorrelation of the signal to non-ideal thermal control of the environment. From Fig.~\ref{fig_reversibility}, one can also see that the correlation may be slightly larger for the cooling stage of the sample than for the heating step. This is probably due to a small temporal delay between the measured temperature and the actual temperature inside the sample. Thus, the above tests clearly prove the reversibility of the material deformation under investigation, and that the experiment will address elastic deformations. Therefore, any effect related to an irreversible deformation or a displacement of the sample may be safely neglected as soon as correlation is smaller than $\sim 0.98$ during the temperature increase. We also checked a great reproducibility of the experiment by observing consistent measurements over several consecutive temperature/wavelength cycles.

\subsection{Correlation functions}

We now consider in more details the correlation functions $g_I(\delta T,\delta \lambda/\lambda)$. These functions may be calculated for every couple of images, but it is very time-consuming and, therefore, we propose several ways to optimise the analysis. First, there are many couples of images corresponding to the same value of  $(\delta T,\delta \lambda/\lambda)$. We verified that for such couples the values of $g_I$ are the same, and in the following we consider only one couple to obtain $g_I(\delta T,\delta \lambda/\lambda)$. Second, one can also see from Eqs.~$\eqref{gE}$-$\eqref{gI}$ that there is a symmetry, such that  $g_I(\delta T,\delta \lambda/\lambda)=g_I(-\delta T,-\delta \lambda/\lambda)$. It allows us to consider only correlation function with $\delta \lambda/\lambda > 0$. In addition, we computed all correlation functions during the increase of temperature. In practice, if $T_{min}$ and $T_{max}$ are the minimum and maximum temperatures, and $\lambda_{min}$ and $\lambda_{max}$ are the minimum and maximum wavelengths respectively, then $g_I(\delta T,\delta \lambda/\lambda)$ is obtained by correlating images of the sample heating at $(T_{min},\lambda_{min})$ and at $(T_{min}+\delta T,\lambda_{min}+\delta \lambda)$ for $\delta T>0$. On the other hand, images recorded at $(T_{min},\lambda_{min}+\delta \lambda)$ and at $(T_{min}-\delta T,\lambda_{min})$ are used to compute the correlation function for $\delta T<0$.

\begin{figure}[htbp]
\centering
\includegraphics[width=1.0\columnwidth]{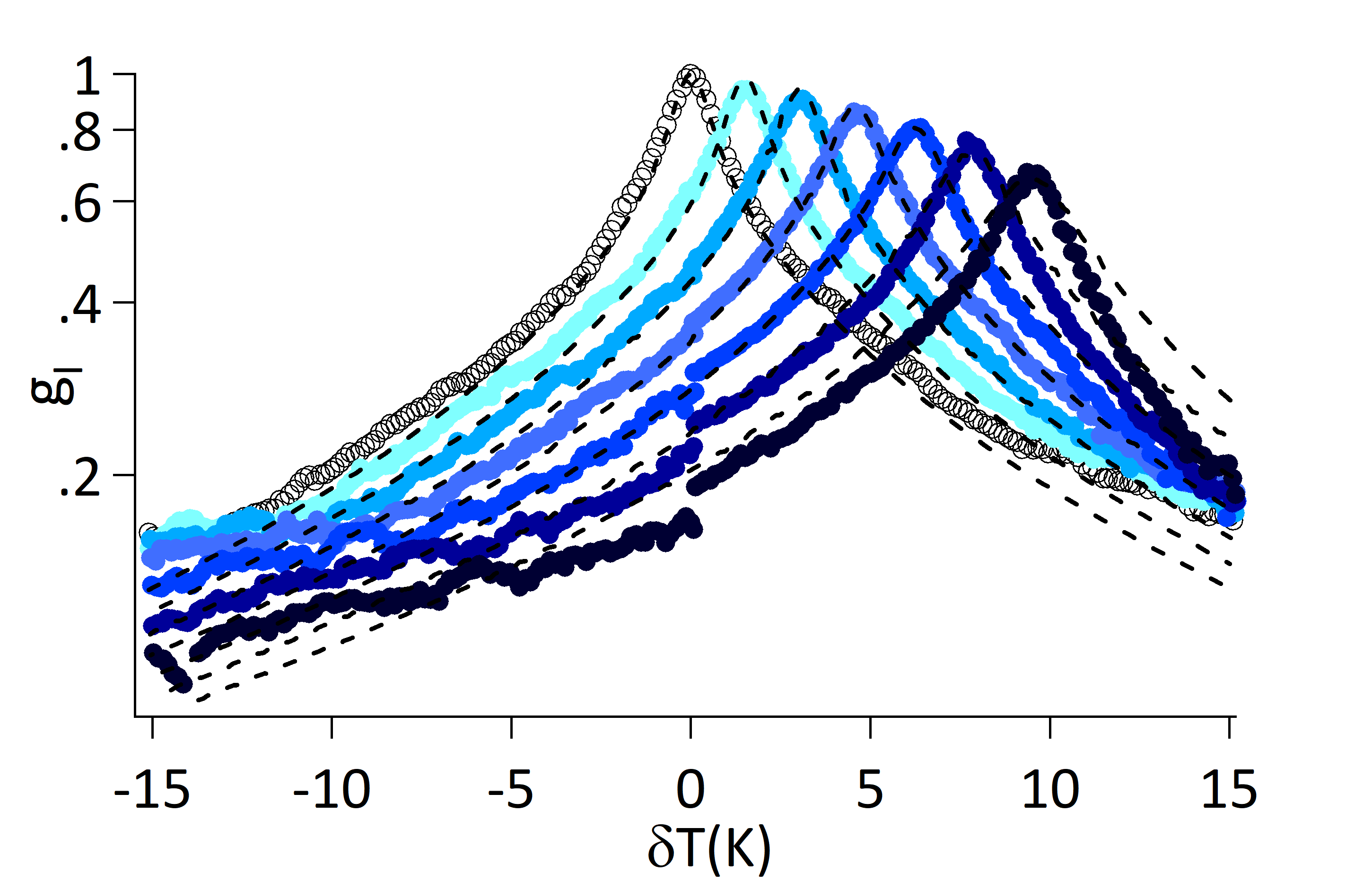}
\caption{Correlation function as a function of $\delta T$. Circle symbols represent the experimental values. Different colors correspond to 7 different values of  $\delta \lambda/\lambda$ ranging from $0$ (white) to $69.7~ 10^{-6}$ (dark) by steps of $11.6~ 10^{-6}$. The dashed lines are the adjustments obtained with the presented model.}
\label{fig_gdtdl}
\end{figure}

The correlation functions obtained for different values of $\delta \lambda/ \lambda$ are shown in Fig.~\ref{fig_gdtdl}. At $\delta\lambda/\lambda=0$, the correlation function logically has a maximum at $\delta T=0$ and it monotonically decreases with temperature variation. It reveals a change in the path lengths induced by thermal expansion of the sample and by variation of the refractive index with the temperature. For a given value of $\delta \lambda/ \lambda$, the correlation function is maximal for a non-vanishing value of $\delta T$ and it monotonically decreases  with temperature around this maximum. It indicates that some part of the path length changes induced by the same reasons are cancelled in this case by wavelength variations. However, the correlation at maximum is lower than one, and decreases with $\delta \lambda / \lambda$.

\subsection{Fitting of the correlation functions.}\label{sec.IV.c}

We now want to describe the correlation function $g_{I}$ in accordance with the model developed in Section~\ref{sec.theory}. The correlation function depends on variables $\delta T$ and $\delta \lambda/ \lambda$, and on numerous parameters. The parameters $A$ and $B$, respectively introduced in Eq.~\eqref{eq_Anu} and \eqref{eq_dphiN_mean2}, are related to the variation of the optical path lengths with temperature. On the other hand, the parameter $C$ introduced in  Eq.~\eqref{eq_dphiN_mean2}, as well as $l^*$, $z_0$, $z_e$ and $\alpha_a$ characterize the propagation of light inside the material. Some of these parameters can be evaluated directly. In the following, we consider a point zone located at the distance of $z_0=l^*$ inside the material. The value of $z_e/l^*$ depends on refractive index of the material. Taking an average refractive index $n_{frit}\simeq 1+\phi_g ~ (n_g-1)\simeq 1.31$, we have~\cite{vera.1996} $z_e=1.8~l^*$. The value of $\alpha_a$ is related to the light absorption in the material, and is of very minor importance in practice in our case. Indeed, the effect of the absorption consist in multiplying $P(s)$ by $\exp^{-s/l_a}$ with $l_a=3/\alpha_a^2 l^*$. However, the finite thickness $L$ of the frit effectively cuts the paths longer than $\sim L^2/l^*$ in $P(s)$. In practice, the better adjustment of the correlation function for $g_I \simeq 1$ is obtained for values of the optical length penetration such that $l_a \gg L^2/l^*$, with no sensitivity of the adjustment to the value of $l_a$. {\jerome The parameter $B$ was fixed to its expected value, evaluated from Eq.~\eqref{eq_B} of Appendix \ref{sec.app.BC.simp}, i.e., $B \simeq 12$. However, in any case, since $B\,A^2\ll C$, it may be seen from Eq.~\eqref{eq_dphiN_mean2} that this parameter does not influence the decay of the correlation function}.

The remaining parameters are then $A$, $C$, and $l^*$. In the case of our model, a fit of the whole dataset from Fig.~\ref{fig_gdtdl} using a chi-square minimization algorithm is difficult. We tried to use a Levenberg-Marquardt algorithm, but this led to convergence problems. More precisely, the algorithm fails to converge, or it converges to parameter values that are very close to initial ones. This can be related either to the strong non-linearity of the $g_I$ function (for $\delta \lambda=0$ and small $\delta T$, $g_I\sim\exp[\sqrt{\delta T}]$, or to the two-dimensional variables for the fitting problem.
Because of all these reasons, we adjusted the parameters manually. This can be easily done since different variables play distinct roles. Indeed, the couple $(\delta T, \delta \lambda /\lambda)$ which maximizes the correlation function depends mainly on the parameter $A$, whereas the  value of the maximum is governed by $C$. Finally, the decay of the correlation function depends on $l^*$. The curves in dashed line in Fig.~\ref{fig_gdtdl} show the modeled correlation functions with the parameters $A_{exp}=7.3\pm 0.1 ~ 10^{-6}$ K$^{-1}$, $l^*=230 \pm 20$ $\mu$m, and $C_{exp}=5 \pm 1.0 ~ 10^{-8} $ K$^{-2}$, where the error bars are subjective.

\section{Discussion}~\label{sec.discussion}

As can be seen from  the adjustment curves provided in Fig.~\ref{fig_gdtdl}, the whole set of correlation functions may be finally correctly described with a limited number of parameters. In this section, we now discuss the values of the adjusted parameters obtained from the DWS experiment. First, the ratio between the estimated transport mean free path $l^*=230 \pm 20~\mu$m and the average pore size of the glass frit sample $d_p=40-100~\mu$m is $\simeq 3$ according to our analysis. Since the typical size of the material is very large with respect to the wavelength, the propagation of light should obey geometrical optics, and the ratio between the transport mean free path and the typical scale of the material should of the order of one as it is observed in foams~\cite{Vera.2001} or granular materials~\cite{djaoui.2005,crassous.2007,mikhailovskaya.2019}. The exact value of this ratio depends on the structure of the material~\cite{gittings.2004} and on the values of refractive indices in the material~\cite{crassous.2007}.

The calculation of parameter $A$ is detailed in Appendix~\ref{sec.app.A}, where we show that
\begin{equation}
A =\alpha+ \frac{\overline{\eta}}{\overline{n}} ~ \Big(\frac{\partial n_g}{\partial T}\Big).
\end{equation}
The quantity $\overline{\eta}$ represents the mean fraction of the length of an optical path which propagates into the glass. Due to internal reflections at glass/air interfaces, this ratio should be slightly larger than the volume fraction of glass $\phi_g$, but less than one. Using conservation of energy in a two-phase system, Gittings {\it et al.}~\cite{gittings.2004} related $\overline{\eta}$ to the volume ratio of the two phases, and to the coefficient of optical reflection and transmission at interfaces. Using this model, we can estimate $\overline{\eta} \simeq 0.72$, and $\overline{n}=1+(n_g-1)\,\overline{\eta}\simeq 1.33$. Taking $\alpha=3.25~ 10^{-6}$ K$^{-1}$, and  $({\partial n_g}/{\partial T})=7.5\pm 2.5 ~ 10^{-6}$ K$^{-1}$ (see Section~\ref{sec.experiment.A}), we obtain $A=7.30\pm 1.35~ 10^{-6}$ K$^{-1}$,  in fair agreement with the value of $A_{exp}$ retrieved experimentally.

The value of $C_{exp}$ deserves a special focus in this work since it encompasses the  information on non-affine deformations. First, we verify that the measured value of $C_{exp}$ cannot be caused by thermo-refractive effects, i.e., variation of the optical index with temperature. It must be noted that even if one assumes a hypothetical situation where non-affine deformations are absent, it is not possible to cancel all available path lengths using a wavelength shift. This is due to the fact that all the light paths are not equally sensitive to thermo-optic variations. The phase variation for a hypothetical path which lies entirely in the air is cancelled when $\delta \lambda/\lambda=\alpha ~ \delta T$, whereas for a path  that would remain entirely inside the glass the cancellation would occur for $\delta \lambda/\lambda=\big(\alpha +({\partial n_g}/{\partial T})\big)~ \delta T$. Since the material deformation leads to a variation of the relative path lengths in the air and in the glass, the exact phase cancellation cannot be obtained simultaneously for all the paths, even when the expansion of the material is strictly affine. In this case the variance of the phase shift varies as $(\delta T)^2$ and it can be derived, as shown in Appendix~\ref{sec.app.deltaC}. Referring to Eq.~\eqref{eq_dphiN_mean2}, the variance of the phase shift due to thermo-refractive effects must yield an additional contribution to the coefficient $C$, which we will denote as $\delta C$ and which we derive in Appendix~\ref{sec.app.deltaC}:
\begin{align}\label{eq:deltaC}
\delta C &=\frac{k \langle l_\nu\rangle  \big(\frac{\partial n_g}{\partial T}\big)^2}{2\overline{n}} ~
\nonumber\\
\Big[\mathcal{F} \frac{\overline{\eta}^2}{\overline{n}^2}(\doverline{n^2}-\overline{n }^2)&+
\mathcal{H}~(\doverline{\eta^2}-\overline{\eta}^2)+
\mathcal{M} \frac{\overline{\eta}}{\overline{n}}(\doverline{\eta n}-\overline{\eta}~\overline{n})\Big].
\end{align}
Quantities such as $\overline{\eta}$, $\overline{n}$, \textit{etc.} are related to the distribution of segment lengths into glass and air. We expect all these quantities to be close to one, and the differences such as $(\doverline{\eta^2}-\overline{\eta}^2)$ to be smaller than 1. Quantities such as $\mathcal{F}$, $\mathcal{H}$ and $\mathcal{M}$ are related to the decay of the correlation function of the segment lengths (see Appendix~\ref{sec.app.deltaC}). We do not expect such correlations on different segments, and then $\mathcal{F},\mathcal{H},\mathcal{M}\simeq 1$, so the term in the brackets should be smaller than 1. To check the validity of this assumption, a numerical model for the light transport in glass spheres (see Appendix~\ref{sec.app.num}) can be used to estimate the value of the bracketted term in Eq.~\eqref{eq:deltaC} to be $0.87$. With this value, and with $\langle l_\nu\rangle =40~\mu m$ and $({\partial n_g}/{\partial T})=7.5~ 10^{-6}$ $K^{-1}$, we obtain $\delta C<0.08 ~ 10^{-8}$ $K^{-2}$ which is $60$ times smaller than the measured value $C_{exp}=5 \pm 1.0 ~ 10^{-8}$ $K^{-2}$. Thus, we conclude that thermo-optic effects cannot explain the lack of correlation observed experimentally.

Once discussed and discarded the potential influence of thermo-optic effects, the proposed geometric model of non-affinity predicts that (see Appendix~\ref{sec.app.BC.simp}):
\begin{equation}
C =\frac{k \langle l_\nu\rangle  \mathcal{G}}{2\overline{n}} ~ \doverline{n^2 (\bfe \cdot \bm{\beta})^2},
\label{eq_C}
\end{equation}
where $\mathcal{G}$ characterizes the auto-correlation of $n_\nu l_\nu (\bfe_\nu \cdot \bm{\beta}_\nu)$ along a path. With $\langle l_\nu\rangle =40~\mu m$, $\overline{n}=1.31$ and taking $n=\overline{n}$ in Eq.~\eqref{eq_C}, we obtain $\mathcal{G} ~ \doverline{(\bfe \cdot \bm{\beta})^2} \approx 4.7~10^{-10}$ $K^{-2}$. This quantity can be normalised to the amount of affine deformation per Kelvin, i.e. $\alpha$, and we finally obtain:
\begin{equation}
\mathcal{G} ~ \doverline{(\bfe \cdot \bm{\beta})^2} \approx 44 ~ \alpha^2.
\end{equation}
The magnitude of this term is apparently quite surprising. Indeed, from the definition of $\bm{\beta}_\nu$ given in Eq.~\eqref{eq_beta_nu}, the ratio of the non-affine displacement $\bm{\beta}_\nu ~ l_\nu ~ \delta T$ to the affine displacement $\alpha ~ l_\nu ~ \delta T$ is:
\begin{equation}
\frac{\bm{\beta}_\nu}{\alpha} \sim \frac{\vert \delta \bfu(\bfr_{\nu+1})-\delta \bfu(\bfr_\nu)\vert}{\vert \bfu(\bfr_{\nu+1})-\bfu(\bfr_\nu)\vert}.
\end{equation}
To our best knowledge, quantitative measurements of elastic non-affine displacements in hyperstatic systems are totally missing for any experimental systems. Some numerical experiments on spring networks such as reported in Ref.~\cite{wyart.2011} give estimates of this ratio (denoted $\delta V^\perp_{n.a.}$ in Ref.~\cite{wyart.2011}), which may be larger to one close to isostaticity, and diminishes in well connected networks~\cite{vanHecke.2009,wyart.2011}. Also, the differences in non-affine deformations have been reported between a system submitted to shear and a system submitted to compression~\cite{tanguy.2002}. As a result, an estimate of the magnitude $\bm{\beta}_\nu/\alpha$ from the data in literature is speculative. However, owing to the fact that we do not study a mechanical system which is close to isostaticity, we expect the ratio $\bm{\beta}_\nu/\alpha$ to be smaller than one.

We now end up this section by discussing the spatial extent of such non-affine deformations in the material. Numerical studies on molecular amorphous systems tend to indicate that the non-affine deformations are correlated over domains of spatial extent $\xi_c$ which are $\sim 10-100$ times larger than the intermolecular microscopic size. Referring to these studies, the characteristic size being $\langle l_\nu\rangle $ in our system, we should expect that $\xi_c \sim  10-100~\langle l_\nu\rangle $. Let us evaluate the value of the factor $\mathcal{G}$ in Eq.~\eqref{eq_C}, which is the sum of the correlation function of the deformation, $g(\nu-\nu')$, appearing the  Eq.~\eqref{eq_g} of Appendix \ref{sec.app.BC.simp} recalled below:
\begin{align}
\big\langle n_\nu l_\nu n_{\nu'} l_{\nu'} &(\bfe_\nu \cdot \bm{\beta}_\nu)(\bfe_{\nu'} \cdot \bm{\beta}_{\nu'})\big\rangle =\nonumber \\
&g(\nu-\nu') \big\langle n_\nu^2 l_\nu^2 (\bfe_\nu \cdot \bm{\beta}_\nu)^2\big\rangle .
\end{align}
 The typical length of the random walk of light in a scattering medium with a transport mean free path $l^*$ inside a non-affine domain of size $\xi_c$ is $\sim \xi_{c}^2/l^*$. This means that the function $g(\nu-\nu')$ should go to $0$ after  $\vert \nu-\nu'\vert \sim \xi_c^2/l^* \langle l_\nu\rangle $ propagation steps of typical length $\langle l_\nu\rangle $, and then $\mathcal{G}$ should be of the order of $\sim \xi_c^2/l^* \langle l_\nu\rangle $. Taking $l^* \simeq 3~ \langle l_\nu\rangle $, we get:
\begin{equation}
\frac{\xi_c}{\langle l_\nu\rangle} ~ \frac{{\doverline{(\bfe \cdot \bm{\beta})^2}}^{1/2}}{\alpha}\sim 11.
\label{eq_xi_c}
\end{equation}
This last equation \eqref{eq_xi_c} indicates that the product of the size of the non-affinity by the relative amount of non-affine deformation is of order of $10$. It is in agreement with a measured relative non-affinity of $10\%$ for $\xi_c/\langle l_\nu\rangle  \sim 100$.

\section{Conclusion}\label{sec:conclu}
We presented an experimental study based on coupled effects between thermal expansion of a material and dilatation of the light wavelength used to probe the material deformation in a DWS setup. We investigated both theoretically and experimentally the relative wavelength variation for which such elastic mechanical deformations may be partially compensated. Such partial compensation was experimentally probed using a coherent scattering experiment and through the analysis of speckle intensity correlation functions. As a result of this study, we were able to measure a coefficient (named $A$) which encompasses the effect of the thermal dilatation and of the thermo-optic expansion, whereas the amount of $-$ imperfect $-$ recorrelation may be interpreted as a deviations of the actual deformation from the affine one. Finally, the proposed model allowed us to quantitatively relate the lack of recorrelation to the amplitude of non-affine deformations.

Although it was not the primary goal of this study, the measure of the coefficient $A$  deserves some metrological interest. Indeed, this experimental method is suitable for measuring relative contractions or expansions in the range of $10^{-6}-10^{-5}$. For this purpose, the value of the thermo-optic coefficient $(\partial n/\partial T)$ is needed, or a calibration  with a sample of known expansion can be also performed. In addition, this method  can provide a measure of  the thermo-optic coefficient of the scattering material, a quantity which is hardly measured with conventional interferometric systems. In that case, the coefficient of linear expansion should be measured by another method.

The main interesting result consists in the analysis of the lack of recorrelation which we related to the occurrence of non-affine deformations in the material. The magnitude of this effect that we observed experimentally is in good agreement with the known orders of magnitude of non-affinity and of the spatial extent $\xi_c$ of domains on which non-affinity is correlated. It is, however, difficult to go further in the comparison between the magnitude of the effect that we measured and the structure of the material. Indeed, theoretical studies on string-connected particles showed that non-affinity depends on the coordination number of the system and on the interacting potential~\cite{ohern.2003}, which are not experimentally available in this experiment. In addition, the presence of an initial stress in the material (frits are prepared from heated glass beads) can have an important impact on the magnitude of non-affinity~\cite{alexander.1998}. A study on materials with a more controlled structure would be useful in this case. Through similar experiments, we observed that such non-affine deformations are also present for a packing of non-connected glass beads. Experiments are slightly more difficult to perform in that case due to irreversible deformations of the materials~\cite{amon2017,djaoui.2005} caused by the difference between the expansion of the granular material and its container. Anyway, the presence of indeterminacy due to solid friction between the grain in contacts make elastic properties of the material dependent on the history of the material preparation~\cite{somfai.2007}. Materials made of sintered or glued beads with a controlled volume fraction and with a measure of connectivity using X-ray tomography may be interesting.

Finally, the presented principle of phase compensation for the scattered wave in order to study the geometry of expansion or contraction of materials can be extended to any kinds of waves where coherent and wavelength-controlled sources are available.

\section{Appendices}

{\jerome \subsection{Influence of the variation of optical path outside the materials}\label{sec.app.outside}

\begin{figure}[htbp]
\centering
\includegraphics[width=.6\columnwidth]{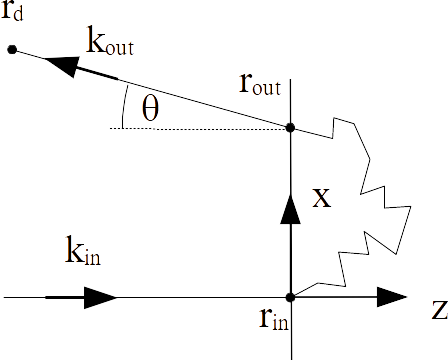}
\caption{Schematic illustration of light propagation near the sample surface close to the laser beam entrance.}
\label{fig1_reply}
\end{figure}

Let us consider a light ray entering into the sample at a point $\bfr_{in}$, exiting at $\bfr_{out}$, detected at point  $\bfr_d$ as sketched in Figure~\ref{fig1_reply}. We denote by $\phi$ the total optical phase accumulated along the path between the source and the detector. For the sake of simplicity, we suppose normal incidence, and we put the origin at $\bfr_{in}$. In addition to the phase shift inside the material, some phase shift occurs due to the displacement of the input, output, and detector points. This additional "external" phase shift is denoted by $\delta \phi_{ext}$, and reads $\delta \phi_{ext}=\bfk_{in}\cdot \delta \bfr_{in}+ \bfk_{out}\cdot(\delta \bfr_{d}-\delta \bfr_{out})$. Under isotropic dilatation, we must have $\delta \bfr_{out}=\alpha~\delta T~(\bfr_{out}-\bfr_0)$ as well as $\delta \bfr_{in}=\alpha~\delta T~(\bfr_{in}-\bfr_0)$, where $\bfr_0$ is some fixed point in the system. Under normal incidence, we have $\bfk_{in} \cdot (\bfr_{out}-\bfr_{in})=0$, and it follows that, after simplification:

\begin{align}
\delta \phi_{ext}&=\bfk_{out}\cdot\delta \bfr_{d}
+\alpha \delta T (\bfk_{out}-\bfk_{in})\cdot\bfr_{0}\nonumber\\
&+\alpha~\delta T~(\bfk_{in}-\bfk_{out})\cdot \bfr_{out}.
\label{deltaphi_ext}
\end{align}

The first term of Eq.~\eqref{deltaphi_ext} is related to the possible displacement of the detector, and the second term corresponds to a global translation of the speckle pattern. Those two terms can be cancelled out with an appropriate choice of $\delta \bfr_{d}$ as it is done experimentally (see Section \ref{sec.experiment.B}). The last term depends on the output point. For a direction of detection near the normal of the sample surface, we have $\bfk_{out}=k~(\theta_x,\theta_y,-1)$ with $\theta_x,\theta_y \ll 1$, and $\delta \phi_{ext} \simeq -k \alpha \delta T~(\theta_x x_{out} + \theta_y y_{out})$. The decorrelation due to this phase shift may be written as :

\begin{align}
g_E=\int & e^{-j k \alpha \delta T\,(\theta_x x_{out} + \theta_y y_{out})}\nonumber\\ &\quad \times P(s,x_{out},y_{out}) ds~dx_{out}~dy_{out}\label{ge.diff.T}
\end{align}

We have introduced the normalized distribution $P(s,x_{out},y_{out})$ of paths with length $s$ exiting at point $(x_{out},y_{out})$. The method for evaluating the integral of Eq.~\eqref{ge.diff.T} is very similar to Ref.~\cite{zhu.1991}. The distribution $P(s,x_{out},y_{out})$ is obtained from the analytical solutions of the diffusion equation. For a diffusing localized source located at $(x=0,y=0,z=l^*)$, we have~\cite{zhu.1991}:
\begin{align}
P(s,x_{out},y_{out})=P(s) \frac{3}{4 \pi s l^*}e^{-3 (x_{out}^2+y_{out}^2)/4sl^*}.
\end{align}
It follows that:
\begin{align}
g_E &=\int P(s) \frac{3}{4 \pi s l^*}
e^{-3 (x_{out}^2+y_{out}^2)/4sl^*}\nonumber\\
&\quad\times e^{j k \alpha \delta T(\theta_x x_{out} + \theta_y y_{out})} ds~dx_{out}~dy_{out}\nonumber\\
&=\int P(s) e^{-\alpha^2 \delta T^2 k^2~(\theta_x^2+\theta_y^2) s l^*/3} ds\nonumber\\
&=\int P(s) e^{p_{ext}s}ds
\end{align}
with
$p_{ext}=-\alpha^2 \delta T^2 k^2~(\theta_x^2+\theta_y^2) l^*/3$.

This amount of decorrelation may be compared to the measured one. For that purpose, we rewrite $p_{ext}$ as $p_{ext}=-k \bar{n} C_{ext} \delta T^2$ with $C_{ext}=\alpha^2 (\theta_x^2+\theta_y^2) k l^* / 3 \bar{n}$. Taking the conservative value $(\theta_x^2+\theta_y^2)^{1/2} \approx 10 \deg$ for the detection angle, we obtain $C_{ext}\simeq 4.10^{-11}K^{-2}$.

This  may be compared to the correlation function of Eq.~\eqref{gE}: $g_E =\int P(s) e^{ps}ds$. At compensation, the temperature and wavelength variations obey $A\delta T=\delta \lambda / \lambda$, and Eq.~\eqref{p} then becomes $p=-k \bar{n} C_{exp} \delta T^2$, with an experimental value of $C_{exp}= 5.10^{-8} K^{-2}$ (see section \ref{sec.IV.c}).

As a result, since $C_{ext}\simeq 4.10^{-11}K^{-2} \ll C_{exp}= 5.10^{-8} K^{-2}$, we may conclude that the variation of phase shift outside the sample is very small compared to the measured one, and can be fairly neglected in the analysis of this experiment.

\subsection{Influence of potential path changes with modification of Fresnel's coefficients}\label{sec.app.B}

Evaluation of correlation functions in Eq.~\eqref{eq:corr_func} implicitly assumes that all the paths before and after deformation are the same. However because refractive indices are changed with temperature and wavelength variations, the reflection and transmission coefficients also change, and some paths may no longer exist after temperature or wavelength variations. To evaluate the potential influence and significance of such an effect, we first consider a path involving $N$ scattering events. Let us denote by $r_i$ the reflection coefficient (in amplitude) at point $\bfr_i$. The probability that a variation of the refractive index modifies a reflection event into a transmission should obviously scale with $| \delta r_i |$, where $\delta r_i$ is the change of reflection coefficient at the interface. As a consequence, in the limit of small $\delta r_i$'s, let us assume that such probability can be evaluated by $|\delta r_i|$.

As a result, for a path involving $N$ scattering events, the probability that the path is unchanged is $p_{id}=\prod_i^N(1-|\delta r_i|)\simeq(1-\langle\delta r\rangle)^N\simeq e^{-N\langle\delta r\rangle}$. We have introduced $\langle\delta r\rangle$ as the absolute value of the reflection coefficient averaged over all orientations of the scattering vectors, and we used the fact that $N\langle\delta r\rangle \ll 1$ in the limit of small changes of refractive indices.

Now, it can be noted that the paths where a reflection becomes a transmission (and conversely) are fully decorrelated, whereas correlation is almost perfectly preserved in the opposite case (for small $\delta r_i$'s). Summing over all path lengths with $s=N\langle l_\nu\rangle$, we obtain:

\begin{align}
g_E&\simeq\int_s P(s) [0.(1-p_{id})+ 1.p_{id}]~ds\nonumber \\
&=\int_s P(s) e^{-s\langle\delta r\rangle/\langle l_\nu\rangle}~ds\nonumber \\
&=\int_s P(s) e^{p_{\delta r}s}~ds,
\end{align}
with $p_{\delta r}=\langle\delta r\rangle/\langle l_\nu\rangle$. The quantity $\langle\delta r\rangle$ may be estimated from Fresnel's formulas of the coefficients of reflection for various incidence angles and polarizations. In the case where the probability of incidence under an angle $\theta$ is $\propto \cos(\theta)$, $\langle\delta r\rangle$ may be computed analytically as a function of the refractive indices of the materials~\cite{duntley.1942}. In our case, we found that $d\langle r\rangle/dn \simeq 0.15$, where the numerical factor $0.15$ arises from the analytical expression of $\langle r\rangle$ for interfaces between media of refractive indices $1$ and $n_g$.

It follows that $p_{\delta r} \simeq 0.15 (\partial n_g / \partial T) \delta T/\langle l_\nu\rangle$. For $\delta T= 10\ K$ and $(\partial n_g/\partial T)=7.5~ 10^{-6}\ K^{-1}$,  we have $p_{\delta r}=0.28~m^{-1}$. This may be compared to the value of $p=k \bar{n} C_{exp} \delta T^2=30~m^{-1}$ found experimentally. In summary, the effect of path changes due to a change of reflection coefficients  at the air/glass interfaces upon temperature variation is small as compared to the measured decorrelation, and can be fairly neglected in the model and the analysis of the results. As noted previously in Section \ref{sec.theory}, the potential influence of wavelength variation is further negligible, and hence does not require additional investigation here.
}

\subsection{Calculus of phenomenological coefficient A}\label{sec.app.A}
From Eq.~\eqref{eq_dphiN_bis} and Eq.~\eqref{eq_Anu} we have :
\begin{align}
\big\langle \delta\phi_N\big\rangle _N&=k N \Bigl[\big\langle n_\nu l_\nu\big\rangle  (-\frac{\delta \lambda}{\lambda}) +\delta T ~ \nonumber \\
&\Bigl[  \big\langle n_\nu l_\nu\big\rangle  \alpha + \big\langle n_\nu l_\nu \bfe_\nu \cdot \bm{\beta}_\nu\big\rangle +\big\langle  \eta_\nu l_\nu \big\rangle  (\frac{\partial n_g}{\partial T})\Bigr].
\label{eq_A1}
\end{align}
First, we do not expect correlation between $l_\nu n_\nu$ and $\bfe_\nu \cdot \bm{\beta}_\nu$, so that $\big\langle n_\nu l_\nu \bfe_\nu \cdot \bm{\beta}_\nu\big\rangle =\big\langle n_\nu l_\nu\big\rangle \big\langle  \bfe_\nu \cdot \bm{\beta}_\nu\big\rangle =0$. Now, using the definition of $\bm{\beta}_\nu$ given in Eq.~(\ref{eq_beta_nu}), and the constraint imposed in Eq.~\eqref{eq_constraint1}, one has $\big\langle \bfe_\nu \cdot \bm{\beta}_\nu\big\rangle =0$. It follows that:
\begin{equation}
A =\alpha+ \frac{\overline{\eta}}{\overline{n}} ~ \big(\frac{\partial n_g}{\partial T}\big).
\label{eq_A}
\end{equation}

\subsection{Coefficient B \& C - Approximate calculus}\label{sec.app.BC.simp}

We first compute coefficients B and C under the simplification that:
\begin{align}
A_\nu&=\alpha +  \frac{\eta_\nu}{n_\nu} ~ \big(\frac{\partial n_g}{\partial T}\big) + \bfe_\nu \cdot \bm{\beta}_\nu\nonumber\\
&\simeq\alpha +  \frac{\overline{\eta}}{\overline{n}}~\big(\frac{\partial n_g}{\partial T}\big) + \bfe_\nu \cdot \bm{\beta}_\nu\nonumber\\
&= A+\bfe_\nu \cdot \bm{\beta}_\nu,
\end{align}
where the last step stems from Eq.~(\ref{eq_A}). In that case, we can rewrite:
\begin{equation}
\delta\phi_N\simeq k \sum_{\nu=0}^{N-1}  l_\nu n_\nu
\big[A \delta T -\frac{\delta \lambda}{\lambda} + \bfe_\nu \cdot \bm{\beta}_\nu \delta T\big].
\label{eq_dphaseN}
\end{equation}
The meaning of this simplification is the following: without non-affinity (i.e. $\bm{\beta}_\nu=0$), the relative variation of the optical path length with temperature is the same for all the paths. Therefore, the path length variations can be cancelled simultaneously everywhere in the sample if $A \delta T=\delta \lambda / \lambda$. In reality, the fraction of the photons' trajectory lying in the glass phase depends on each path. Thus, the cancellation is not possible, even when  non-affinity is absent. This contribution to the parameter $C$ is discussed in the next section of the Appendix. Yet, using the above simplification, we obtain:
\begin{align}
\langle \delta & \phi_N^2\rangle _N=k^2 \Big\langle  \sum_{\nu,\nu'} n_\nu l_\nu n_{\nu'} l_{\nu'}~ \Big[ (A \delta T -\frac{\delta \lambda}{\lambda})^2
\nonumber\\
&+(\bfe_\nu \cdot \bm{\beta}_\nu)(\bfe_{\nu'} \cdot \bm{\beta}_{\nu'})~ (\delta T)^2+(\text{cross~terms}) \Big] \Big\rangle ,
\end{align}
where $\nu$ and $\nu'$ vary between $0$ and $N-1$. Under the same assumptions of decorrelation between the non-affinity and the optical transport, the cross terms are null. The average of $n_\nu l_\nu n_{\nu'} l_{\nu'}$ must depend only on $\nu-\nu'$, then:
\begin{align}
\big\langle n_\nu l_\nu n_{\nu'} l_{\nu'}\big\rangle &=f(\nu-\nu') \big\langle n_\nu^2 l_\nu^2\big\rangle \nonumber \\
&+ (1-f(\nu-\nu'))\big\langle n_\nu l_\nu\big\rangle ^2,\label{eq_f}
\end{align}
where $f$ is the unknown even discrete autocorrelation function of $n_\nu l_\nu$ when the latter quantity is considered as a discrete function defined over segment indices $\nu\in[0; N-1]$. Such autocorrelation function verifies $f(0)=1$ and $f(|\nu-\nu'|)\to 0$ when $\vert \nu-\nu' \vert \to \infty$. See works of Bicout {\it et al.}~\cite{bicout.1991,bicout.1993} for the use of correlation functions along photon random walks, and the Appendix~\ref{sec.app.num} for an example of numerical computation of $f$. If we set the "integral" (discrete summation) of the autocorrelation function as $\mathcal{F} = \sum_{\nu-\nu'=-\infty}^{+\infty}f(\nu-\nu')$, then we can write:
\begin{align}
\Big\langle  \sum_{\nu,\nu'} n_\nu l_\nu n_{\nu'} l_{\nu'} \Big\rangle &-
\Big\langle  \sum_{\nu} n_\nu l_\nu  \Big\rangle ^2 =\nonumber \\
&N ~ \mathcal{F} ~ \Big[\big\langle  n_\nu^2 l_\nu^2\big\rangle -\big\langle n_\nu l_\nu\big\rangle ^2\Big],
\end{align}
where we assumed that the decay of the autocorrelation function $f$ is rapid compared to $N$. This is a reasonable assumption as long-range correlation across path length segments is not expected.

Similarly we can write:
\begin{align}
\big\langle n_\nu l_\nu n_{\nu'} l_{\nu'} &(\bfe_\nu \cdot \bm{\beta}_\nu)(\bfe_{\nu'} \cdot \bm{\beta}_{\nu'})\big\rangle =\nonumber \\
&g(\nu-\nu') \big\langle n_\nu^2 l_\nu^2 (\bfe_\nu \cdot \bm{\beta}_\nu)^2\big\rangle ,
\label{eq_g}
\end{align}
where we took into account again the fact that $\big\langle n_\nu l_\nu (\bfe_\nu \cdot \bm{\beta}_\nu)\big\rangle =0$, and where we introduced the correlation function $g(\nu-\nu')$. With $\mathcal{G} = \sum_{\nu-\nu'=-\infty}^{+\infty}g(\nu-\nu')$, one gets
\begin{align}
\Big\langle  \sum_{\nu,\nu'} n_\nu l_\nu n_{\nu'} l_{\nu'} & (\bfe_\nu \cdot \bm{\beta}_\nu)(\bfe_{\nu'} \cdot \bm{\beta}_{\nu'}) \Big\rangle =\nonumber \\
&N ~ \mathcal{G} ~ \big\langle  n_\nu^2 l_\nu^2 (\bfe_\nu \cdot \bm{\beta}_\nu)^2\big\rangle .
\end{align}

Finally,
\begin{align}
\langle \delta \phi_N^2\rangle _N&-\langle \delta \phi_N\rangle _N^2=N  k^2 ~[\nonumber\\
\Big[&\mathcal{F}~ \big(\big\langle  n_\nu^2 l_\nu^2\big\rangle -\big\langle n_\nu l_\nu\big\rangle ^2\big) ~ \big(A \delta T -\frac{\delta \lambda}{\lambda}\big)^2\nonumber\\
+&\mathcal{G}~ \big\langle  n_\nu^2 l_\nu^2 (\bfe_\nu \cdot \bm{\beta}_\nu)^2\big\rangle  ~ (\delta T)^2\Big]
\end{align}
so that, from Eq.~\eqref{eq_dphiN_mean2}, one gets:
\begin{equation}
B =\frac{k \langle l_\nu\rangle \mathcal{F}}{2\overline{n}} ~ \bigl[\doverline{n^2}-\overline{n}^2\bigr]
\label{eq_B}
\end{equation}
and
\begin{equation}
C =\frac{k \langle l_\nu\rangle \mathcal{G}}{2\overline{n}} ~ \doverline{n^2 (\bfe \cdot \bm{\beta})^2}.
\label{eq_C_annexe}
\end{equation}

\subsection{Coefficient C - Correction $\delta C$ due to the thermo-refractive coefficient.}\label{sec.app.deltaC}

To derive the expression of coefficients B and C, we previously made the simplifying assumption that $\eta_\nu/n_\nu \simeq  \overline{\eta}/\overline{n}$. If we relax this approximation, the phase shift becomes:
\begin{align}
\delta\phi_N = k \sum_{\nu=0}^{N-1}  l_\nu n_\nu
\Big[A \delta T -\frac{\delta \lambda}{\lambda} + \bfe_\nu \cdot \bm{\beta}_\nu \delta T\nonumber \\
+\big(\frac{\eta_\nu}{n_\nu}-\frac{\overline{\eta}}{\overline{n}}\big) \big(\frac{\partial n_g}{\partial T}\big) \delta T \Big].
\label{eq_dphin_complete}
\end{align}
We now need to evaluate the effect of the last term of the above equation on the variance of $\delta \phi_N$. For the sake of simplicity, we evaluate this term close to the correlation recovery $A\delta T = \delta \lambda /\lambda$, and without non-affinity of deformation (i.e., $\bfe_\nu \cdot \bm{\beta}_\nu=0$), so that:
\begin{align}
\delta\phi_N = k \big(\frac{\partial n_g}{\partial T}\big) \delta T\sum_{\nu=0}^{N-1}
\big(l_\nu \eta_\nu - l_\nu n_\nu \frac{\overline{\eta}}{\overline{n}}).
\end{align}

Introducing the autocorrelation function of $\eta_\nu l_\nu$:
\begin{align}
\big\langle \eta_\nu l_\nu \eta_{\nu'} l_{\nu'}\big\rangle &=h(\nu-\nu') \big\langle \eta_\nu^2 l_\nu^2\big\rangle \nonumber \\
&+ (1-h(\nu-\nu'))\big\langle \eta_\nu l_\nu\big\rangle ^2,
\label{eq_h}
\end{align}
and the cross-correlation function between $\eta_\nu l_\nu$ and $n_\nu l_\nu$:
\begin{align}
\big\langle \eta_\nu l_\nu n_{\nu'} l_{\nu'}\big\rangle &=m(\nu-\nu') \big\langle \eta_\nu n_\nu l_\nu^2\big\rangle \nonumber \\
&+ (1-m(\nu-\nu'))\big\langle n_\nu l_\nu\big\rangle  \big\langle \eta_\nu l_\nu\big\rangle ,
\label{eq_m}
\end{align}
and setting their respective sums as $\mathcal{H} = \sum_{\nu-\nu'=-\infty}^{+\infty}h(\nu-\nu')$ and $\mathcal{M} = \sum_{\nu-\nu'=-\infty}^{+\infty}m(\nu-\nu')$, we obtain:
\begin{align}
&\langle \delta \phi_N^2\rangle _N-\langle \delta \phi_N\rangle _N^2=N  k^2 \big(\frac{\partial n_g}{\partial T}\big)^2 (\delta T)^2~\nonumber\\
&\Big[\mathcal{F} \frac{\overline{\eta}^2}{\overline{n}^2}(\doverline{n^2}-\overline{n }^2)+
\mathcal{H}~(\doverline{\eta^2}-\overline{\eta}^2)+
\mathcal{M} \frac{\overline{\eta}}{\overline{n}}(\doverline{\eta n}-\overline{\eta}~\overline{n})\Big].
\end{align}

From the definition of $B$ and $C$ in Eq.~\eqref{eq_dphiN_mean2}, the above equation indicates that the corrective term
\begin{align}
\delta C &=\frac{k \langle l_\nu\rangle  \big(\frac{\partial n_g}{\partial T}\big)^2}{2\overline{n}} ~
\nonumber\\
\Big[\mathcal{F} \frac{\overline{\eta}^2}{\overline{n}^2}(\doverline{n^2}-\overline{n }^2)&+
\mathcal{H}~(\doverline{\eta^2}-\overline{\eta}^2)+
\mathcal{M} \frac{\overline{\eta}}{\overline{n}}(\doverline{\eta n}-\overline{\eta}~\overline{n})\Big]
\label{eq_C'}
\end{align}
must be added to Eq.~\eqref{eq_C_annexe}.

\subsection{Numerical estimate of correlation functions $f$, $h$, $m$ and their sums.}\label{sec.app.num}

The geometrical characteristic of the optical paths in the material can be estimated using a model of ray propagation inside a dense packing of spheres previously developed in Ref.~\cite{crassous.2007}. We consider a random close packing of identical spheres (density $\phi=0.637$, radius $R$, refractive index $n_g=1.456$). For every segment, we obtain $l_\nu$, $\eta_\nu$ and $n_\nu$, and we can compute correlations functions such as $\big \langle  n_\nu l_\nu n_{\nu'} l_{\nu'} \big\rangle $.

\begin{figure}[htbp]
\centering
\includegraphics[width=1.0\columnwidth]{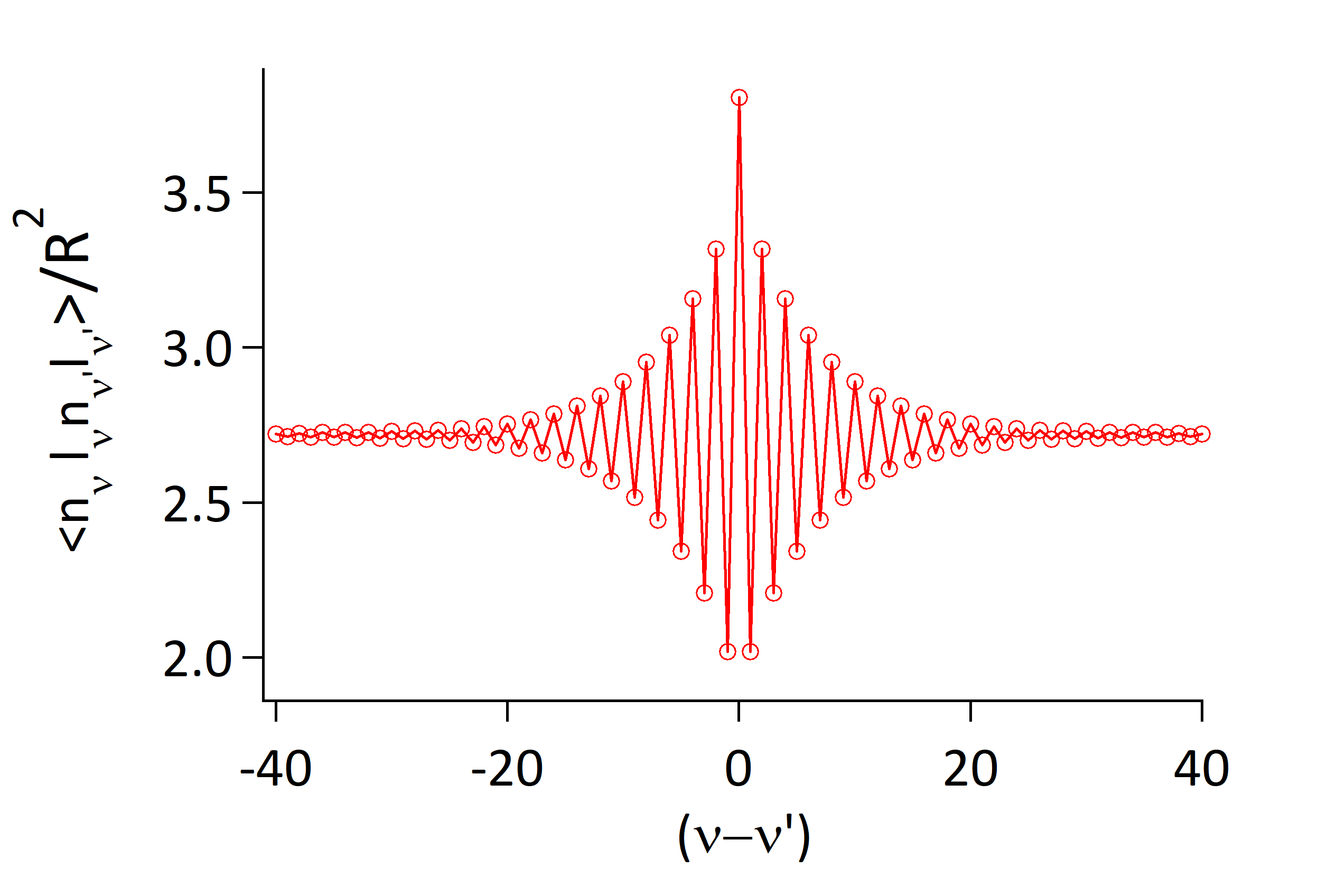}
\caption{Discrete correlation function $\big \langle  n_\nu l_\nu n_{\nu'} l_{\nu'} \big\rangle $ in $R^2$ units as a function of $(\nu-\nu')$.}
\label{fig_c_nl}
\end{figure}

Figure~\ref{fig_c_nl} shows the dependence of the correlation function $\big \langle  n_\nu l_\nu n_{\nu'} l_{\nu'} \big\rangle $ as a function of  $(\nu-\nu')$, with very similar behaviors for $\big \langle  \eta_\nu l_\nu \eta_{\nu'} l_{\nu'} \big\rangle $ and $\big \langle  n_\nu l_\nu \eta_{\nu'} l_{\nu'} \big\rangle $. It can be seen that the correlation function behaves as expected from Eq.~\eqref{eq_f}. The oscillations at small $|\nu-\nu'|$ are due to the fact that the segments in vacuum (with small values of $n_\nu l_\nu$) alternate with the segments in glass (large values of $n_\nu l_\nu$).

From those correlation function along paths of the rays, we can measure (in $R$ units) :

$\big \langle l_\nu \big \rangle =1.25$; $\big \langle l_\nu^2 \big \rangle =2.09$; $\big \langle n_\nu l_\nu \big \rangle =1.65$; $\big \langle n_\nu^2 l_\nu^2 \big \rangle =3.81$; $\big \langle \eta_\nu l_\nu \big \rangle =0.87$; $\big \langle \eta_\nu^2 l_\nu^2 \big \rangle =1.53$; $\big \langle n_\nu \eta_\nu l_\nu^2 \big \rangle =2.22$; $\mathcal{F}~ (\big \langle n_\nu^2 l_\nu^2 \big \rangle -\big \langle n_\nu l_\nu \big \rangle ^2)=0.34$; $\mathcal{H}~ (\big \langle \eta_\nu^2 l_\nu^2 \big \rangle -\big \langle \eta_\nu l_\nu \big \rangle ^2)=0.085$; $\mathcal{M}~ (\big \langle \eta_\nu n_\nu l_\nu^2 \big \rangle -\big \langle n_\nu l_\nu \big \rangle \big\langle \eta_\nu l_\nu \big \rangle )=0.091$.
The bracketted term in Eq.~\eqref{eq_C'} is then $0.23$, whereas $\overline{\eta}=0.70$ and $\overline{n}=1.32$.
We stress the fact that the packing used for this numerical estimation is slightly different from the glass frit that we used in the experiment, with possible difference in the distribution of segment sizes. However, we expect  the deviation from the numerical estimation to be very small.

\begin{acknowledgments}
We acknowledge the funding from Agence Nationale de la Recherche Grant No.ANR-16-CE30-0022. The authors thank Ana\"el Lema\^\i tre for scientific discussions.
\end{acknowledgments}

\end{document}